\documentclass[12pt]{article}
\usepackage{epsfig}
\usepackage{a4wide}

\usepackage{amssymb}
\newcommand{\be}{\begin{equation}}
\newcommand{\ee}{\end{equation}}

\newcommand{\bea}{\begin{eqnarray}}
\newcommand{\eea}{\end{eqnarray}}

\newcommand{\eps}{\varepsilon}
\newcommand{\sla}{a\kern-5pt\raise1pt\hbox{$\scriptstyle/$}\kern2pt}
\newcommand{\slk}{/\kern-6pt k}
\newcommand{\sls}{s\kern-5pt/}
\newcommand{\slt}{\tau\kern-5pt/}
\newcommand{\Tr}{\mathop{\rm Tr}\nolimits}

\begin{document}
\thispagestyle{empty}
\begin{flushright}
MZ-TH/10-29\\
1008.0917 [hep-ph]\\
August 2010
\end{flushright}

\begin{center}
{\Large\bf Symmetries and similarities for spin orientation parameters
 in $e^+e^-\to ZH,Z\gamma,ZZ$ at SM thresholds}\\[24pt]
{\large S.~Groote$^{1,2}$, H.~Liivat$^1$ and I.~Ots$^1$}\\[12pt]
$^1$ Loodus- ja Tehnoloogiateaduskond, F\"u\"usika Instituut,\\
  Tartu \"Ulikool, Riia~142, 51014 Tartu, Estonia\\[12pt]
$^2$ Institut f\"ur Physik der Johannes-Gutenberg-Universit\"at,\\
  Staudinger Weg 7, 55099 Mainz, Germany
\vspace{12pt}
\end{center}

PACS numbers:
13.88.+e, 
12.15.Ji, 
13.66.Bc 

\begin{abstract}
We consider the spin orientation of the final $Z$ bosons for the processes in
the Standard Model. We demonstrate that at the threshold energies of these
processes the analytical expressions for the $Z$ boson polarization vectors
and alignment tensors coincide ($e^+e^-\to ZH,Z\gamma$) or are very similar
($e^+e^-\to ZZ$). In addition, we present interesting symmetry properties for
the spin orientation parameters.
\end{abstract}

\newpage

\section{Introduction}
The Standard Model (SM) of elementary particle physics has been
phenomenologically very successful. However, as a theory it is less impressive.
There are many fundamental questions that remain unanswered by the SM. Due to
this, there are strong reasons to expect that there is a great deal of (new)
physics beyond the SM, more fundamental with a characteristic mass scale. The
present good agreement between the accelerator-based experimental data and the
SM predictions suggests that the energy scale associated with any new physics
model should be much higher than the energy reach of the colliders up to now.
It is believed that the new accelerator, the Large Hadron Collider (LHC),
started in 2009, is able to produce new physics particles, first of all new
on-shell resonances or a single heavy new particle in association with the
SM ones~\cite{Lykken:2010mc}. However, such a scenario for visualizing new
physics experimentally is obviously beyond the reach of future $e^+e^-$
colliders, like the planned International Linear Collider (ILC).

There are methods to probe new physics at energies below the new physics mass
scale. These more indirect scenarios are based on observations of small
deviations from the SM predictions for processes with SM particles where new
physics effects can arise only from non-standard interactions. The price to
pay for such a possibility is the need for higher sensitivity, both
experimentally and theoretically. In the case of future ILC-type colliders,
the achievable sensitivities are expected to be dramatically (i.e.\ up to
orders of magnitude) improved as compared to the ones obtained up to now.
Among the various ways to increase the sensitivity we emphasize two. First,
there is a reason to expect that processes including heavy particles are more
sensitive to new physics manifestations. Second, processes including
polarization effects have higher sensitivity than unpolarized processes. The
possibility at ILC to use both longitudinally and transversely polarized
initial beams provide additional means to increase the sensitivity for new
physics.

Aforementioned circumstances considerably increases the potential to probe
anomalous interactions via spin effects in $e^+e^-$ annihilation into heavy
particles. Therefore, the processes $e^+e^-\to ZH$, $Z\gamma$, and $ZZ$
have been studied extensively. At the same time it is obvious that the search
for new physics, even though performed at very sensitive future colliders,
cannot be successful without knowing the SM predictions with appropriate
precision. In Ref.~\cite{Lykken:2010mc} Lykken writes: ``[\dots] to first
approximation [\dots] experimenters do not need to know anything about BSM
[Beyond the SM] models in order to make discoveries -- but they need to know a
lot of the Standard Model physics!'' In the context of searching for new
physics in aforementioned processes via spin orientation, this means that the
knowledge of all possible spin effects in these processes in the framework of
the SM forms a good basis for rejecting or limiting anomalous couplings. In
this paper we try to add our contribution to this basis.

In the framework of the SM we present and analyse analytical expressions for
the $Z$ boson spin polarization vectors and alignment tensors for the processes
$e^+e^-\to ZH$, $Z\gamma$, and $ZZ$ near the production threshold. For all
these processes the initial beams are taken to have both longitudinal and
transverse polarization components. We demonstrate that the expressions for
the spin orientation parameters in these processes have similar shape and
symmetry properties. Even more, the expressions for the $Z$ boson polarization
vector and alignment tensor for the processes $e^+e^-\to Z\gamma$ and $ZH$
coincide.

The paper is organized as follows. In Sec.~2 we explain how spin orientation
parameters can be described by density matrices. In addition we show how to
obtain the actual spin polarization vector and alignment tensor of the final
$Z$ boson. In Sec.~3 we present the analytical expressions for the spin
orientation parameters of the three processes and dwell on similarities and
interesting symmetry features. The results are checked with the help of
positivity conditions. At the end of this paper we give our conclusions.

\section{Description of spin orientation parameters}
In physical processes the particles with non-zero spins are generally in a
mixed spin state. Contrary to the pure state which can be described by a
single wave function, the description of a mixed state deserves an incoherent
superposition of $2s+1$ orthogonal pure state wave functions. By definition
this superposition means that in order to calculate the probability for finding
a certain mixed state, one has to calculate the average of the probabilities of
the pure states, assigning to each pure state an appropriate weight.

The most natural way to describe mixed states is to use the spin density matrix
formalism. Technically, if interested in the spin orientation of a certain
particle in the process, one replaces the element $u\bar u$ (spin-1/2 case) or
$\eps_\mu\eps_\nu^*$ (spin-1 case) in the squared amplitude of the process by
the mixed state spin density matrix. For our calculations the relativistic
spin-1/2 density matrices for the initial beams and the non-relativistic
spin-1 density matrices for the final $Z$-bosons are needed. The relativistic
spin-1/2 density matrices have the familiar shape
\begin{equation}\label{rhoh}
\rho_\mp=\frac12(\slk_\mp\pm m)(1+\gamma_5\sls_\mp)
\end{equation}
where the upper signs refer to the electron and the lower ones to the positron.
Here and in the following we use Feynman's slash notation
$\sla=a_\mu\gamma^\mu$. The polarization four-vectors are given by
\begin{equation}
s_\mp=(s_0,\vec s_\mp)=\left(\frac{\vec k_\mp\cdot\vec\xi_\mp}m,
  \vec\xi_\mp+\frac{(\vec k_\mp\cdot\vec\xi_\mp)\vec k_\mp}{m(k_0-m)}\right)
\end{equation}
where $\vec\xi_\mp$ are the polarization vectors in the resp.\ rest frame of
the particle. For the absolute value $|\vec\xi_\mp|$ of the mixed state
polarization vectors one can take any value between zero and one. If the
electron and positron have both longitudinal polarization (LP) and transverse
polarization (TP), it is useful to divide the polarization vector $s_\mp$ up
into LP and TP parts. The limit $m/k_0\to 0$ which is used in our calculations
can be conveniently taken by making use of the approximation
\begin{equation}\label{LTP}
s_\mp^\mu\approx\frac{h_\mp k_\mp^\mu}m+\tau_\mp^\mu
\end{equation}
and subsequently setting $m=0$. $h_\mp$ is the measure of LP and
$\tau_\mp=(0,\vec\tau_\mp)$ is the TP four-vector with $\vec\tau_\mp$ the
transverse component of the polarization vector
($\vec k_\mp\cdot\vec\tau_\mp=0$). After inserting Eq.~(\ref{LTP}) into
Eq.~(\ref{rhoh}) and performing the limit $m\to 0$, one obtains~\cite{DassRoss}
\begin{equation}
\rho=\frac12(1\pm h_\mp\gamma_5+\gamma_5\slt_\mp)\slk_\mp
\end{equation}
convenient for our calculations.

For describing mixed states by $3\times 3$ non-relativistic hermitian
spin-density matrixes, at most 8 real parameters are needed. For expanding such
a matrix one can use a basis containing the unit matrix $I$, the three spin
matrices $S_i$, and the six combinations of products of two spin matrices in
the form
\begin{equation}
S_{ij}=\frac32\left(S_iS_j+S_jS_i-\frac43\delta_{ij}\right).
\end{equation}
In such a basis the basic elements are hermitian zero-trace matrices which
makes their usage convenient. The density matrix expanded in this basis reads
\begin{equation}
\rho=\frac13\left(I+\frac32t_iS_i+\frac13t_{ij}S_{ij}\right)
\end{equation}
where here and in the following we sum over all indices which occur twice
(Einstein's convention). Note that this basis is over-determined. Instead of 9
matrices needed to expand the density matrix it contains 10 matrices. However,
not all the elements of the basis are linearly independent. The relation
\begin{equation}
S_{xx}+S_{yy}+S_{zz}=0
\end{equation}
holds for three of the basic elements. In order to reduce the number of
parameters from 9 to 8 independent ones, it is necessary to claim
\begin{equation}\label{traceless}
t_{xx}+t_{yy}+t_{zz}=0.
\end{equation}
Due to this constraint, one obtains
\begin{equation}\label{rhoij}
\rho_{ij}=\frac13\delta_{ij}+\frac12t_k(S_k)_{ij}
  +\frac16t_{kl}(S_kS_l+S_lS_k)_{ij}
\end{equation}
where $t_k=\Tr(\rho S_k)$ is the polarization vector and
$t_{kl}=\Tr(\rho S_{kl})$ is the orientation or alignment tensor describing
the alignment of spins.

When taking the spin-1 matrices in the representation
$(S_k)_{ij}=-i\epsilon_{ijk}$ where $\epsilon_{ijk}$ is the totally
antisymmetric tensor, one obtains
\begin{equation}
(S_kS_l+S_lS_k)_{ij}=2\delta_{kl}\delta_{ij}-\delta_{ki}\delta_{lj}
  -\delta_{kj}\delta_{li}.
\end{equation}
In this case, $\rho_{ij}$ from Eq.~(\ref{rhoij}) can be written in the
simple form
\begin{equation}
\rho_{ij}=\frac13\left(\delta_{ij}-\frac32it_k\epsilon_{ijk}-t_{ij}\right).
\end{equation}
This form is used in our calculations.

\subsection{\label{Sec21}Spin orientation of final particles}
In our calculations it is supposed that the spin orientation of a single final
$Z$ boson is observed.\footnote{The spin orientation parameters $t_i$ and
$t_{ij}$ determine the angular distribution of the decay products. Therefore,
the $Z$ boson spin orientation parameters can be seen by analyzing the angular
distribution of its decay products. According to the Particle Data
Group~\cite{PDG}, the $Z$ boson decays with $10\%$ probability into lepton
pairs ($e^+e^-$, $\mu^+\mu^-$, $\tau^+\tau^-$), and with $70\%$ probability
into hadron pairs. Considering only lepton pairs which can be clearly
distinguished from the background, the analysis can be found in Sec.~4 of
Ref.~\cite{Ots:2000pq}, giving the correlation between the spin orientation
parameters and the angular distribution.} Accordingly, the squared amplitude
of the process can be given in the form
\begin{equation}\label{M2}
|{\cal M}|^2\sim S+V_it_i+T_{ij}t_{ij}
\end{equation}
where $S$, $V_i$ and $T_{ij}$ are the scalar, vector and tensor built from the
polarization parameters of the initial electron and positron, the kinematical
parameters of all the particles participating in the process, and the coupling
constants. Since the calculations are performed in the CM frame with vanishing
electron mass at threshold energies, there are only two kinematical parameters
-- the electron three-momentum $\vec k$ and the mass $M_Z$ of the $Z$ boson.

The squared amplitude determines the probability that the process produces a
final $Z$ boson with spin orientation parameters $t_i$ and $t_{ij}$. On the
other hand, the same probability can be expressed also as the traced product of
two density matrices
\begin{equation}\label{rhorrho}
\Tr(\rho^r\rho)\sim 1+\frac32 t_i^rt_i+\frac13 t_{ij}^r t_{ij}
\end{equation}
where the real (actual) density matrices and their parameters are denoted by
an upper index $r$. The symbols without this index are the density matrices and
their parameters that are substituted into the squared amplitude instead of
$\eps_\mu^Z\eps_\nu^{Z*}$. By comparing Eqs.~(\ref{M2}) and~(\ref{rhorrho})
one can find the actual polarization vector and alignment tensor of the
$Z$ boson,
\begin{equation}
t_i^r=\frac2{3S}V_i,\qquad t_{ij}^r=\frac3S\left(T_{ij}+T\delta_{ij}\right)
\end{equation}
where the term $T$ incorporates the tracelessness~(\ref{traceless}).

\subsection{\label{Sec22}Positivity conditions}
The values of the parameters $t_i$ and $t_{ij}$ are restricted by their
physical boundaries as extremal values. In this physical region parameters are
linear independent. However, due to the positivity (non-negativity) condition
of the density matrices they depend on each other non-linearly. It can be shown
that the most restrictive requirements for the spin-1 density matrix can be
given by the inequality~\cite{Minnaert:1966zz}
\begin{equation}
2\Tr\rho^3-3\Tr\rho^2+1\ge 0.
\end{equation}
If one expresses the density matrix through its parameters, one obtains
\begin{equation}\label{posex}
\frac29-\frac12t_it_i+\frac12t_it_jt_{ij}-\frac19t_{ij}t_{ij}
  -\frac2{27}t_{ij}t_{jk}t_{ki}\ge 0.
\end{equation}
From this expression one can clearly see how the vector and tensor parameters
depend non-linearly on each other. Eq.~(\ref{posex}) is useful when analyzing
the possibilities of tuning the final $Z$ boson spin orientation by varying
the polarization of the initial beams.

\section{$Z$ boson spin orientation parameters at threshold}
We are now ready to present the final $Z$ boson spin orientation parameters
(polarization vectors and alignment tensors) at the threshold energies for the
three processes $e^+e^-\to ZH,Z\gamma,ZZ$, calculated in the center-of-mass
system. Note that the kinematics simplifies significantly when calculating at
the threshold. On the other hand, the results obtained are also valid close
to the threshold.

\begin{figure}[ht]\begin{center}
\epsfig{figure=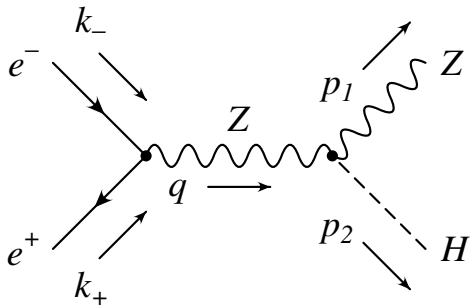, scale=0.5}
\caption{\label{fig1}$s$-channel SM tree level diagram for $e^+e^-\to ZH$ with
  point-like $ZHZ$ vertex}
\end{center}\end{figure}

The first process $e^+e^-\to ZH$ could be important in clarifying the
mechanisms of electroweak symmetry breaking proposed by various beyond-the-SM
models (for an independent work on this subject see e.g.\
Refs.~\cite{Barger:1993wt,Miller:2001bi}). In the SM at tree level this
process is described by an $s$-channel Feynman diagram with a point-like $ZHZ$
vertex (Fig.~\ref{fig1}).
Actually, at lowest order there are three Feynman diagrams corresponding to
the process. However, the two diagrams where the Higgs boson couples to the
electron line can be ignored due to the smallness of the coupling which is
proportional to the electron mass.

\begin{figure}[ht]\begin{center}
\epsfig{figure=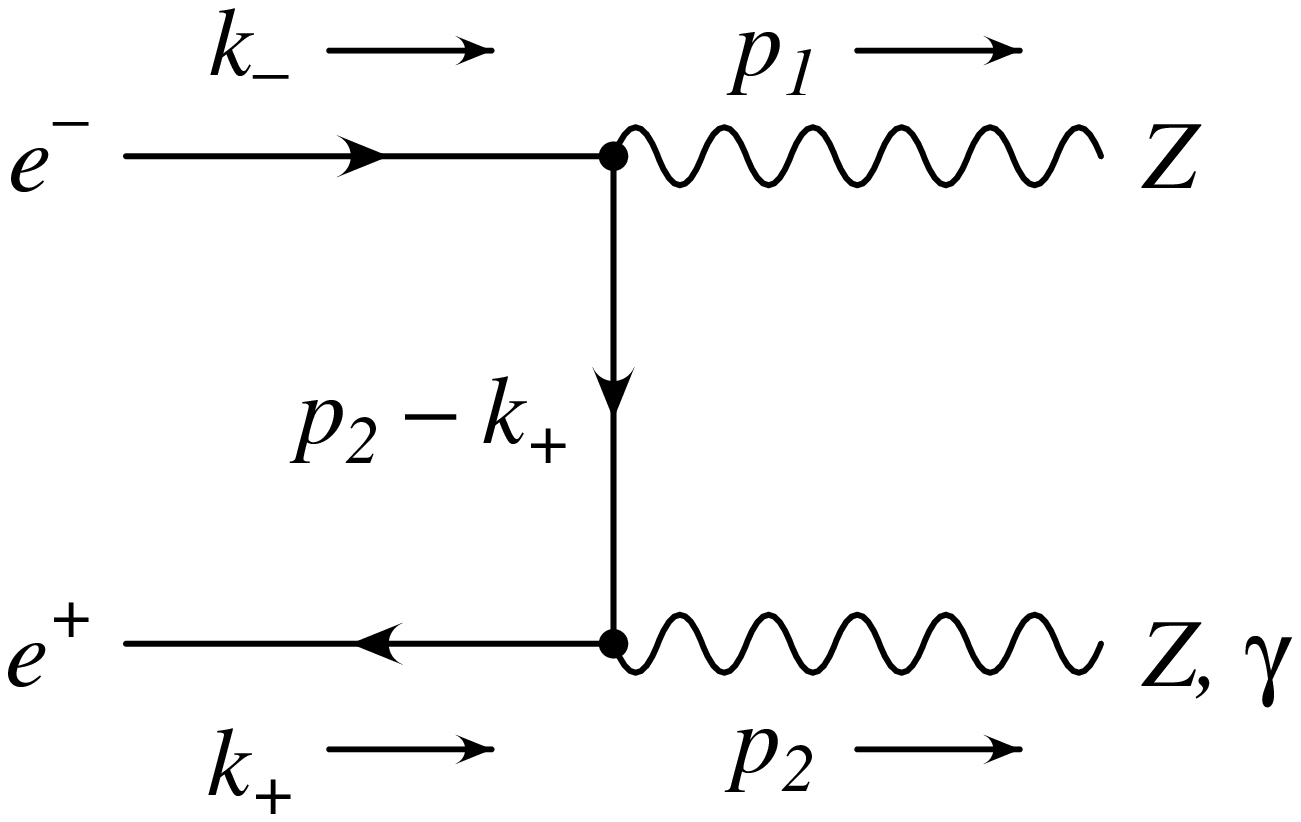, scale=0.5}\qquad
\epsfig{figure=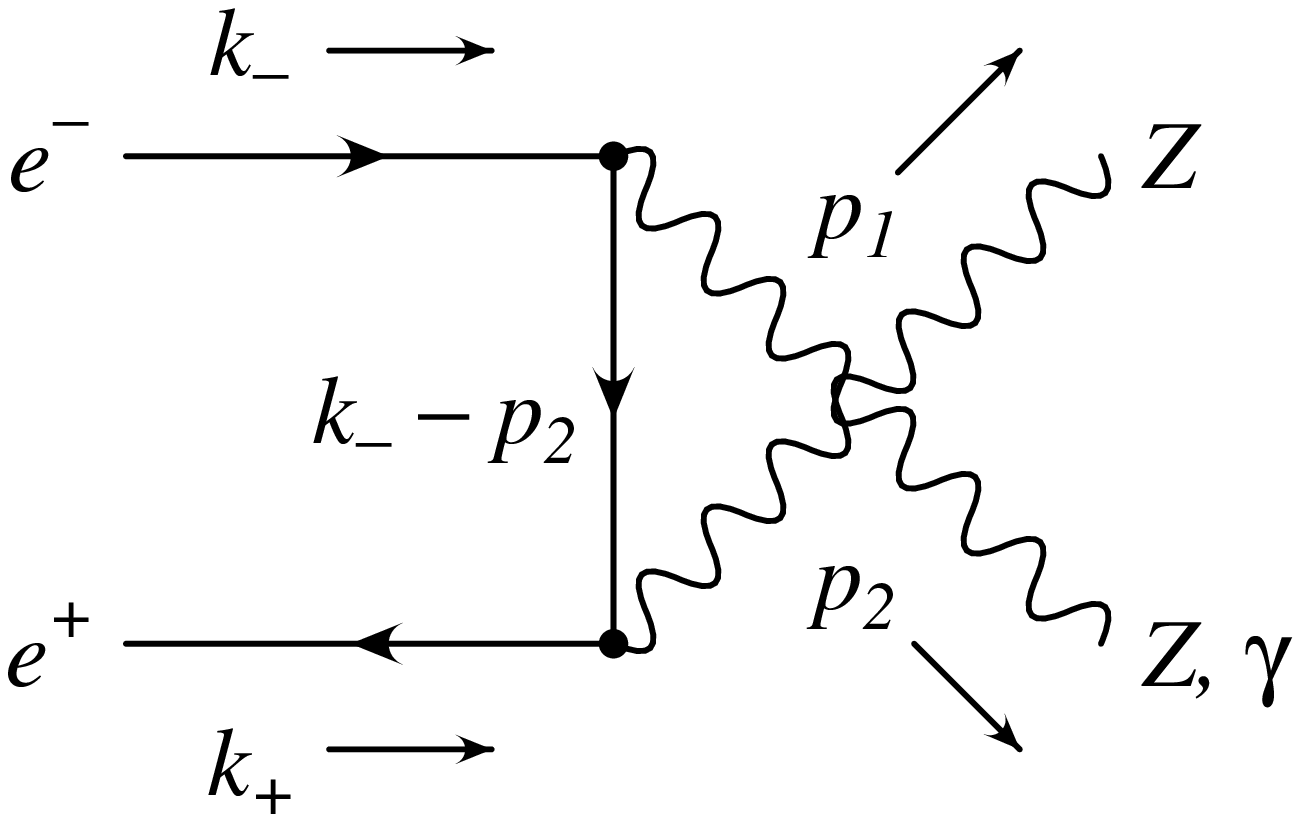, scale=0.5}\\(a)\kern196pt(b)
\caption{\label{fig2}(a) $t$-channel and (b) $u$-channel SM tree level diagram
  for $e^+e^-\to Z\gamma,ZZ$}
\end{center}\end{figure}

The two remaining processes $e^+e^-\to Z\gamma,ZZ$ have been most of all used
to investigate possible new physics manifestations via anomalous gauge boson
self-couplings $Z\gamma\gamma$, $Z\gamma Z$, and $ZZZ$. In the SM at tree level
these processes are described by a $t$-channel and an $u$-channel diagram
(Fig.~\ref{fig2}).

The $Z$ boson spin polarization vectors and alignment tensors for the processes
$e^+e^-\to ZH,Z\gamma$ with vanishing electron mass have been calculated by us
earlier~\cite{Ots:2000pq,Ots:2004hk}. At the threshold energies the spin
orientation parameters reduce to the same expressions,
\begin{eqnarray}
\vec t^{\,Z}_{\rm thres}&=&\frac{f_-^{(2)}}{f_+^{(2)}}\hat k,\label{tiZ}\\
t^Z_{ij\,{\rm thres}}&=&\frac32(\hat k_i\hat k_j-\frac13\delta_{ij})
  +\frac{3g_Lg_R}{f_+^{(2)}}\Big[(\vec\tau_-\cdot\vec\tau_+)(\hat k_i\hat k_j
  -\delta_{ij})+\tau_{-i}\tau_{+j}+\tau_{-j}\tau_{+i}\Big]\label{tijZ}
\end{eqnarray}
where we used $g_L=\frac12(g_V+g_A)$, $g_R=\frac12(g_V-g_A)$ in order to write
the expressions in a more symmetric form, and the short hand notation
$f_\pm^{(2)}:=g_L^2(1-h_-)(1+h_+)\pm g_R^2(1+h_-)(1-h_+)$. $h_\mp$ and
$\vec\tau_\mp$ are the measures of the electron (positron) longitudinal
polarizations and the electron (positron) transverse polarization vectors,
respectively. $\hat k$ indicates the unit vector along the electron momentum.

Even though somewhat peculiar, the coincidence of the $Z$ boson spin
orientation parameters of a $s$-channel diagram with the spin orientation
parameters of a $t$-channel and an $u$-channel diagram at corresponding
thresholds is expected. One can say that Eqs.~(\ref{tiZ}) and~(\ref{tijZ})
give the spin orientation parameters of the real $Z$ boson in the process
$e^+e^-\to Z$. The Higgs boson is a spin-0 particle and, therefore, cannot
affect the spin orientation of the $Z$ boson at the threshold of
$e^+e^-\to ZH$. In the same way, for $\vec p_1\to 0$ the photon in the process
$e^+e^-\to Z\gamma$ can be considered as a radiative correction to the main
process $e^+e^-\to Z$. The coincidence of the $Z$ boson orientation parameters
for the two different processes, therefore, can be taken as a cross-check for
the final expressions.

In order to obtain the spin orientation parameters for one of the final $Z$
bosons in the process $e^+e^-\to ZZ$ at threshold, we again take the (massless)
electron and positron beams to be polarized, having both longitudinal and
transverse polarization components. In summing over the spin orientations of
the second $Z$ boson (in Fig.~\ref{fig2} indicated with momentum $p_2$), we
obtain for the squared amplitude of the process at threshold
\begin{eqnarray}\label{MZZ}
\lefteqn{|{\cal M}^{ZZ}_{\rm thres}|^2\ =\ \frac{32e^4}{3\sin^2(2\theta_W)}
  \Bigg\{f_+^{(4)}+\frac34f_-^{(4)}\hat k\cdot\vec t
  -\frac14f_+^{(4)}\hat k_i\hat k_jt_{ij}\strut}\nonumber\\&&\strut
  +\frac{g_L^2g_R^2}4\Big[(\vec\tau_-\cdot\vec\tau_+)\hat k_i\hat k_j
  +\tau_{-i}\tau_{+j}+\tau_{-j}\tau_{+i}\Big]t_{ij}\Bigg\}
\end{eqnarray}
where the short-hand notation
$f_\pm^{(4)}:=g_L^4(1-h_-)(1+h_+)\pm g_R^4(1+h_-)(1-h_+)$ is used. According
to the rules given in Sec.~\ref{Sec21}, for the spin orientation parameters one
obtains
\begin{eqnarray}
\vec t^{\,ZZ}_{\rm thres}&=&\frac{f_-^{(4)}}{2f_+^{(4)}}\hat k,\label{tiZZ}\\
t^{ZZ}_{ij\,{\rm thres}}&=&-\frac34(\hat k_i\hat k_j-\frac13\delta_{ij})
  +\frac{3g_L^2g_R^2}{2f_+^{(4)}}
  \Big[(\vec\tau_-\cdot\vec\tau_+)(\hat k_i\hat k_j-\delta_{ij})
  +\tau_{-i}\tau_{+j}+\tau_{-j}\tau_{+i}\Big].\label{tijZZ}
\end{eqnarray}
Note that the terms $\delta_{ij}$ are introduced into $t_{ij}$ in order to
make it traceless. Obviously, the spin orientation parameters in
Eqs.~(\ref{tiZZ}) and~(\ref{tijZZ}) do not coincide with those of the processes
$e^+e^-\to ZH,Z\gamma$ (Eqs.~(\ref{tiZ}) and~(\ref{tijZ})) but are similar to
them. This is quite understandable. Due to the two $Z$ boson final state the
constants $g_L$ and $g_R$ appear twice in the squared amplitude. Therefore,
the squares in $f_\pm^{(2)}$ had to be replaced by fourth powers in
$f_\pm^{(4)}$. On the other hand, only the direction $\hat k$ of the electron
and the transverse polarization vectors $\vec\tau_\pm$ are available to build
the spin orientation parameters. As a result, besides the change of the power
of the coupling constants the expressions can differ only by the constants in
front of the various three contributions. This is indeed the case in the
expressions obtained.

\subsection{Symmetry properties of the polarization vectors}
In this section we demonstrate that the polarization vectors of the $Z$ boson
in the different processes under consideration feature interesting symmetries.
Expressing the polarization vectors as a function of the effective polarization
parameter
\begin{equation}
\chi=\frac{h_+-h_-}{1-h_+h_-}
\end{equation}
of the initial beams, for the processes $e^+e^-\to ZH,Z\gamma$ we obtain
$\vec t^{\,Z}_{\rm thres}=t^Z(\chi)\hat k$ where
\begin{equation}
t^Z(\chi)=\frac{\chi+a}{a\chi+1},\qquad
  a=\frac{g_L^2-g_R^2}{g_L^2+g_R^2}=\frac{2g_Vg_A}{g_V^2+g_A^2}\approx 0.147
\end{equation}
where we used $g_V=\frac12(-1+2\sin^2\theta_W)\approx-0.037$,
$g_A=-\frac12=-0.5$, $g_L=-\frac12\cos^2\theta_W$, and
$g_R=\frac12\sin^2\theta_W$ (with $\sin^2\theta_W=0.2315$,
$\cos^2\theta_W=0.7685$~\cite{PDG}). For the process $e^+e^-\to ZZ$ one obtains
$\vec t^{\,ZZ}_{\rm thres}=t^{ZZ}(\chi)\hat k$ where
\begin{eqnarray}
t^{ZZ}(\chi)&=&\frac12\,\frac{\chi+b}{b\chi+1}\hat k,\nonumber\\
b&=&\frac{g_L^4-g_R^4}{g_L^4+g_R^4}\ =\ \frac{4g_Vg_A(g_V^2+g_A^2)}{(g_V^2
  +g_A^2)^2+4g_V^2g_A^2}\ \approx\ 0.294.
\end{eqnarray}
For longitudinally unpolarized initial beams ($\chi=0$) one obtains quite
similar values $t^Z(\chi=0)=a\approx 0.147$ and
$t^{ZZ}(\chi=0)=\frac12b\approx 0.144$. The similarity is related to the
smallness of $g_V$. Exact coincidence would hold for $g_V=0$. However, in this
case one would obtain $a=b=0$.

On the other hand, the polarizations vanish for $\chi=-a$ and $\chi=-b$,
respectively, i.e.\ $t^Z(\chi=-a)=0$ and $t^{ZZ}(\chi=-b)=0$. This is because
the reciprocal functions read
\begin{equation}
\chi^Z(t)=\frac{-t+a}{at-1},\qquad
\chi^{ZZ}(t)=\frac{-2t+b}{2bt-1}.
\end{equation}
Note that for the process $e^+e^-\to t\bar t$ at threshold the SM result for
the final top quark polarization can be expressed as $\xi=\xi(\chi)\hat k$
where the function $\xi(\chi)=-(\chi+c)/(c\chi+1)$ has similar symmetry
properties with $\xi(\chi=0)=\chi(\xi=0)=-c\approx-0.408$~\cite{Groote:2009dd}.

\subsection{Positivity condition for longitudinally polarized initial beams}
\begin{figure}\begin{center}
\epsfig{figure=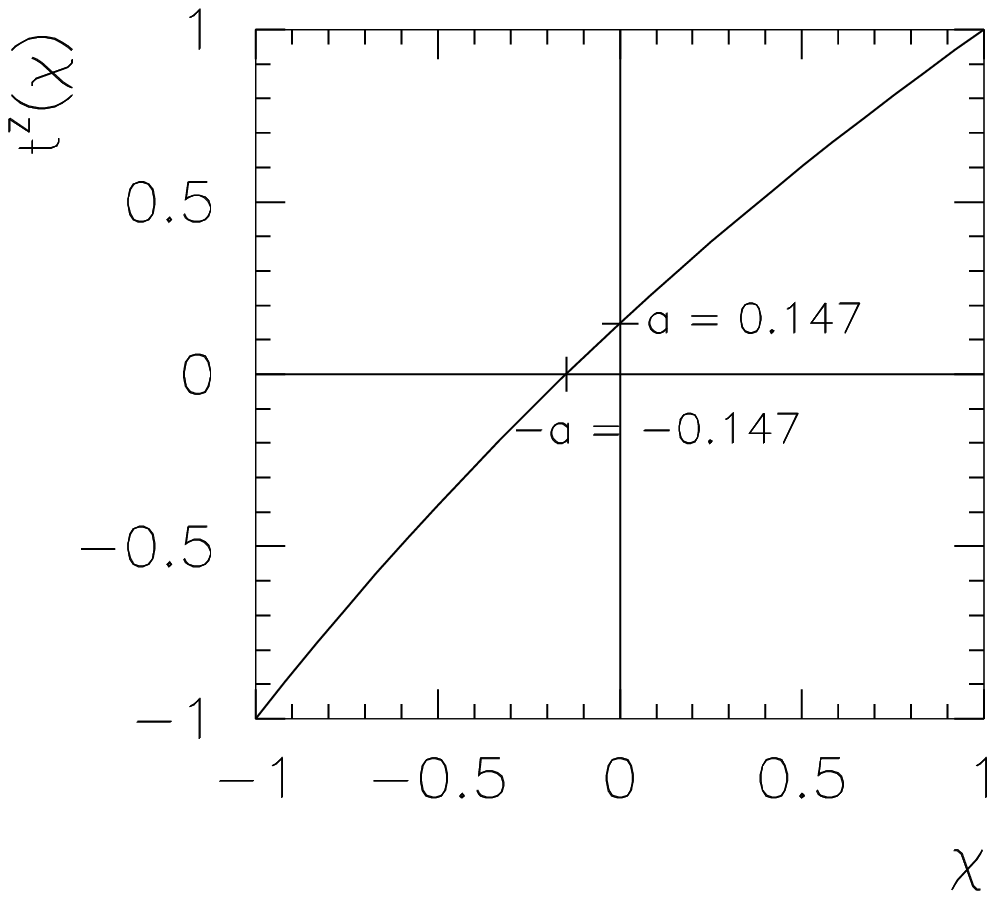, scale=0.7}\qquad
\epsfig{figure=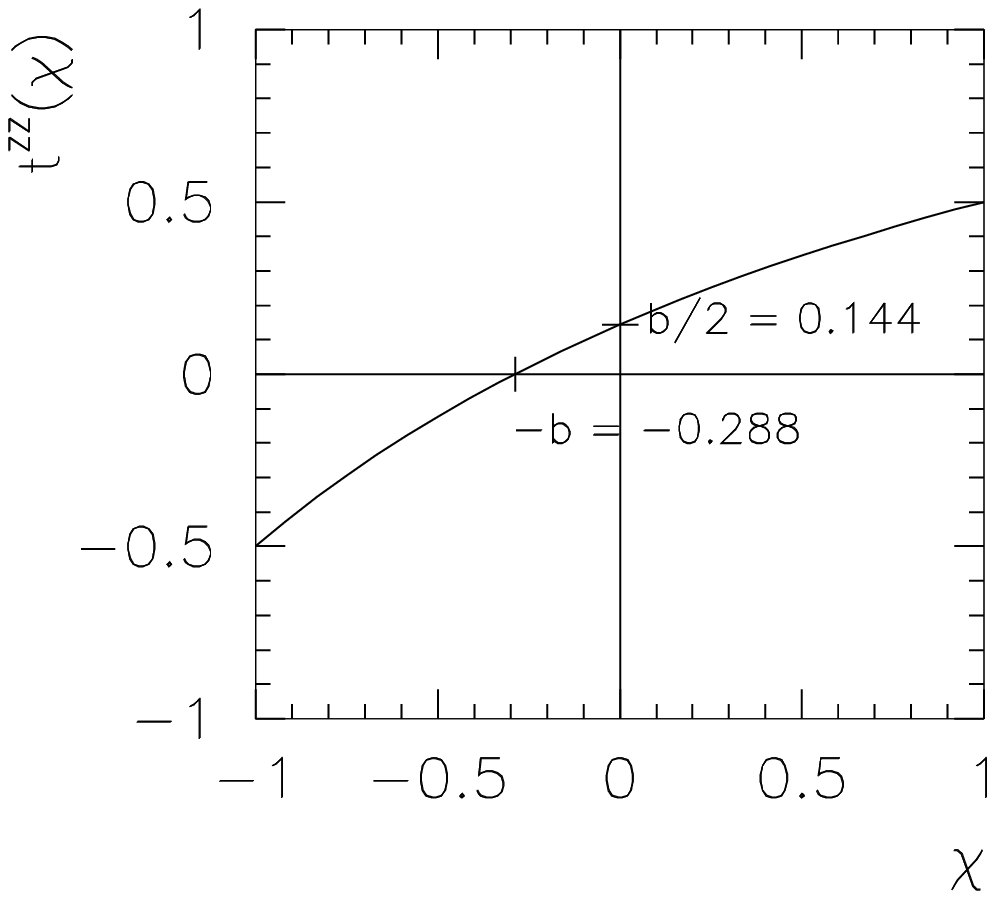, scale=0.7}\\(a)\kern196pt(b)
\caption{\label{fig3}Dependence of (a) $t^Z(\chi)$ and (b) $t^{ZZ}(\chi)$
on the effective polarization parameter $\chi$ of the initial beams}
\end{center}\end{figure}
In Fig.~\ref{fig3} the dependence of $t^Z$ and $t^{ZZ}$ as functions of the
effective polarization parameter $\chi$ is presented. Note that $\chi$ can
obtain values between $-1$ and $+1$ only if the initial beams have no
transverse polarization components. While $t^Z$ takes all values between
$-1$ and $+1$, the range for $t^{ZZ}$ is restricted to $|t^{ZZ}|\le \frac12$.
This is mirrored also by the positivity condition in Sec.~\ref{Sec22}, giving
a possibility  to check our result in an independent way.

As mentioned before, the polarization vectors and the alignment tensors can
depend only on the direction $\hat k$ of the electron beam and the transverse
polarization vectors of the initial beams. Considering first longitudinally
polarized beams ($\vec\tau_\pm=0$) and choosing the $z$-axis to be the
direction of the electron beam, one obtains the non-vanishing components
\begin{eqnarray}
t^Z_z&=&\frac{\chi+a}{a\chi+1}\hat k_z\ =\ \frac{\chi+a}{a\chi+1},\nonumber\\
t^Z_{zz}&=&1,\qquad t^Z_{xx}\ =\ t^Z_{yy}\ =\ -\frac12.
\end{eqnarray}
Inserting the $t_{ij}$ into the positivity condition~(\ref{posex}) it turns
out that the condition is satisfied independently of the value of $t^Z_z$ and,
therefore, also independently of the value of $\chi$. Hence, by varying the
effective polarization parameter $\chi$ one can theoretically force the $Z$
boson polarization vector to take any value between $-1$ and $+1$. Practically
achievable values are not far from the theoretical ones. When using the values
$h_-=\pm0.8$, $h_+=\pm0.6$, planned to be achieved at the ILC, one can reach
value $\chi=\pm0.95$ for the effective polarization parameter, leading to
similar values for the $Z$ polarization (cf.\ Fig.~\ref{fig3}(a)).

When doing the same with the components of the spin orientation parameters for
the process $e^+e^-\to ZZ$, one obtains
\begin{eqnarray}
t^{ZZ}_z&=&\frac12\,\frac{\chi+b}{b\chi+1}\hat k_z
  \ =\ \frac12\,\frac{\chi+b}{b\chi+1},\nonumber\\
t^{ZZ}_{zz}&=&-\frac12,\qquad t^{ZZ}_{xx}\ =\ t^{ZZ}_{yy}\ =\ \frac14.
\end{eqnarray}
Inserting the $t_{ij}$ into the positivity condition~(\ref{posex}),
one obtains $|t^{ZZ}_z|\le 1/2$ which is consistent with Fig.~\ref{fig3}(b).

\subsection{Positivity condition as a test for the result}
Also in the general case, i.e.\ with transversely and/or longitudinally
polarized initial beams, the positivity condition~(\ref{posex}) is in harmony
with the spin orientation parameters $t_i$ and $t_{ij}$ of the processes
$e^+e^-\to ZH,Z\gamma,ZZ$. Our investigation checks the obtained expressions
for the spin orientation parameters and shows how much one can tune these
parameters when varying the initial beam polarization parameters $h_\mp$ and
$\vec\tau_\mp$.

Taking both transverse polarization vectors $\vec\tau_-$ and $\vec\tau_+$ to
be directed along the $x$-axis of the coordinate system, for the
processes $e^+e^-\to ZH,Z\gamma$ one obtains
\begin{eqnarray}
t^Z_z&=&\frac{f_-^{(2)}}{f_+^{(2)}},\nonumber\\
t^Z_{xx}&=&-\frac12+\frac{3g_Lg_R}{f_+^{(2)}}\tau_{-x}\tau_{+x},\nonumber\\
t^Z_{yy}&=&-\frac12-\frac{3g_Lg_R}{f_+^{(2)}}\tau_{-x}\tau_{+x},\nonumber\\
t^Z_{zz}&=&1.
\end{eqnarray}
Inserting the $t_{ij}$ into the positivity condition~(\ref{posex}),
one finds that the condition takes the form $0=0$. This means that the
positivity condition is satisfied in any case, i.e.\ for any values $h_\mp$
and $\tau_{\mp x}$. As a consequence, by varying the LP parameters $h_\mp$
(or $\chi$) one can tune the $Z$ boson polarization over the whole range
$[-1,1]$ if $\vec\tau_-$ or $\vec\tau_+$ or both of them are equal to zero.
In the same way, by varying $\vec\tau_\mp$ one can tune the alignment tensor
for $h_-=h_+=0$.

Under the same conditions, for the process $e^+e^-\to ZZ$ one obtains
\begin{eqnarray}
t^{ZZ}_z&=&\frac12\,\frac{f_-^{(4)}}{f_+^{(4)}},\nonumber\\
t^{ZZ}_{xx}&=&\frac14+\frac{3g_L^2g_R^2}{2f_+^{(4)}}\tau_{-x}\tau_{+x},
  \nonumber\\
t^{ZZ}_{yy}&=&\frac14-\frac{3g_L^2g_R^2}{2f_+^{(4)}}\tau_{-x}\tau_{+x},
  \nonumber\\
t^{ZZ}_{zz}&=&-\frac12.
\end{eqnarray}
After inserting $t_{ij}$ into the positivity condition~(\ref{posex}), one
obtains the restriction
\begin{equation}
(t_{ZZ}^z)^2\le\frac14-\frac{g_L^4g_R^4}{f_+^{(4)}}(\tau_{-x}\tau_{+x})^2.
\end{equation}
In absence of the term depending on $\tau_{-x}\tau_{+x}$ this result
reproduces the previous result, i.e.\ that by varying the parameters $h_\mp$
(or $\chi$) one can theoretically tune the $Z$ boson polarization from zero
to $1/2$ ($t^{ZZ}_z=0$ for $\chi=-b=-0.288$ and $t^{ZZ}_z=1/2$ for $\chi=1$).
Finally, if we insert also the value $t_{ZZ}^z$ into the positivity
condition~(\ref{posex}), we obtain a condition of the form
\begin{equation}
(\tau_{-x}\tau_{+x})^2\le(1-h_-^2)(1-h_+^2).
\end{equation}
For $h_-=h_+=0$ the condition $(\tau_{-x}\tau_{+x})^2\le 1$ puts no constraints
on the transverse polarization of the initial beams. On the other hand, if one
of the initial beams is longitudinally polarized, restrictions on the
transverse polarizations of the initial beams are imposed.


\section{Conclusions}
We have presented the spin orientation parameters of the final $Z$ boson for
the three processes $e^+e^-\to ZH$, $Z\gamma$ and $ZZ$ in the Standard Model
at the threshold energies and have found their properties. Note that a part of
the properties found are valid also at higher energies. Therefore, the
processes depend on the longitudinal polarization of the initial beams via
the factors $f_\pm^{(2)}=g_L^2(1-h_-)(1+h_+)\pm g_R^2(1+h_-)(1-h_+)$ or
$f_\pm^{(4)}=g_L^4(1-h_-)(1+h_+)\pm g_R^4(1+h_-)(1-h_+)$ not only at the
thresholds of the processes but also in the general case. Such properties could
be helpful in separating SM contributions from new physics ones.

\subsection*{Acknowledgements}
The work is supported by the Estonian target financed projects No.~0180013s07
and No.~0180056s09. S.G.\ acknowledges support by the Deutsche
Forschungsgemeinschaft (DFG) under grant 436 EST 17/1/06.

\end{document}